\def\kms{\mbox{$\rm km\;s^{-1}$}}
\def\degr{\mbox{$^\circ$}}
\def\ha{\mbox{H$\alpha$}}
\def\nii{\mbox{[N~{\scriptsize II}]}}
\def\msun{\mbox{M$_\odot$}}

\documentstyle[11pt,aaspp4]{article}
\slugcomment{{\it submitted to The Astrophysical Journal Letters 
    on June 25, 1998}}

\lefthead{Bertola et al.}
\righthead{Circumnuclear Keplerian Disks in Galaxies}

\begin{document}

\title{Circumnuclear Keplerian Disks in Galaxies\footnote{Based on
observations carried out at ESO, La Silla, (Chile) (ESO N. 58, A-0564),
and at the Mt. Graham International Observatory (Arizona) with the VATT:
the Alice P. Lennon Telescope and the Thomas J. Bannan Astrophysics
Facility.}}

\author{F.~Bertola\altaffilmark{2}, M.~Cappellari\altaffilmark{2},
J.G.~Funes,~S.J.\altaffilmark{2},\\ E.M.~Corsini\altaffilmark{2},
A.~Pizzella\altaffilmark{3} \& J.C.~Vega~Beltr\'an\altaffilmark{4}}

\altaffiltext{2}{Dipartimento di Astronomia, Universit\`a di Padova,
Vicolo dell'Osservatorio 5, I-35122 Padova, Italy}
\altaffiltext{3}{European Southern Observatory, Alonso de Cordova 3107,
Casilla 19001, Santiago 10, Chile}
\altaffiltext{4}{Telescopio Nazionale Galileo, Osservatorio Astronomico
di Padova, Vicolo dell'Os\-ser\-vatorio 5, I-35122 Padova, Italy}

\begin{abstract}

In this paper we demonstrate the possibility of inferring the presence
of Keplerian gaseous disks using optical ground-based telescopes
properly equipped.

We have modeled the peculiar bidimensional shape of the emission lines
in a sample of five S0-Sa galaxies as due to the motion of a gaseous
disk rotating in the combined potential of a central point-like mass and
of an extended stellar disk.  The value of the central mass
concentration estimated for four galaxies of the sample (NGC~2179,
NGC~4343, NGC~4435 and NGC~4459) is $\sim10^9$ M$_\odot$. For the
remaining galaxy NGC~5064 an upper limit of $5\times10^{7}$ M$_\odot$ is
estimated.

\end{abstract}


\clearpage
\section{Introduction}

There is an increasing evidence of conspicuous mass concentrations in
the center of galaxies, lending support to the idea that their central
engine is constituted by a black hole (see \cite{kor95}, \cite{ho98} for
recent reviews).  This evidence comes both from stellar and gaseous
dynamics. In this latter case the mass concentration is deduced by the
observation of an increase towards the center of the rotation velocity
of the gaseous disk, according to the Kepler's third law.  Seven
circumnuclear Keplerian disks (hereafter CNKD) have been up to now
observed: four have been discovered in elliptical galaxies using the
high resolution capability of the {\it Hubble Space Telescope (HST)\/}
(\cite{fer96,mac97,bow98,vdm98}), while three were detected in spirals
with Very Long Baseline Interferometry (VLBI) and Very Long Baseline
Array (VLBA) observations of maser sources (\cite{miy95,gre97b,gre97c}).

In a program aimed to study with high spatial and spectral resolution
the structure of the emission lines in the nuclear regions of early-type
disk galaxies, we have obtained with the 3.6-m telescope at La Silla
major axis spectra of the Sa galaxies NGC~2179 and NGC~5064. In the
first case it is possible to recognize, applying our modeling technique,
the kinematical behavior typical of a CNKD while the second represents a
limiting case for its detection with our instrumental setup. In addition
we apply our modeling to three early-type disk galaxies (NGC~4343,
NGC~4435 and NGC~4459) observed by Rubin, Kenney \& Young (1997),
showing that also in these cases the shape of the emission lines is
consistent with the presence of CNKDs. In this way we demonstrate the
feasibility of detecting CNKDs with optical ground-based telescopes. The
general data on the galaxies studied are given in Table~1.

\section{Observations and data reduction}
 
The spectroscopic observations of NGC~2179 and NGC~5064 were carried out
at the 3.6-m ESO Telescope in La Silla on February 3-4, 1997. The
telescope was equipped with the Cassegrain Echelle Spectrograph mounting
the Long Camera in long-slit configuration without the crossdisperser.
The 31.6 lines mm$^{-1}$ grating was used in combination with a
1\farcs3$\times$2\farcm4 slit. The spectral order \#86 ($\lambda_c =
6617$ \AA) corresponding to the redshifted \ha\ region was isolated by
means of the narrow-band 6630/51 \AA\ filter. It yielded a wavelength
coverage of about 78 \AA\ between 6593 \AA\ and 6670 \AA\ with a
reciprocal dispersion of 3.17 $\rm \AA\;mm^{-1}$. The adopted detector
was the No.  37 1024$\times$1024 TK1024AB CCD with a 24$\times$24
$\mu$m$^2$ pixel size. No on-chip binning was applied and each pixel
corresponds to $\rm 0.076\,\AA$$\times$0\farcs33.

We took for NGC~2179 and NGC~5064 six and four separate major-axis
spectra (P.A. = 170\degr\ and P.A. = 38\degr\ respectively) centered on
the nucleus for a total exposure time of 360 and 240 minutes
respectively. The galaxies were centered on the slit using the guiding
camera at the beginning of each exposure.  Comparison thorium-argon lamp
exposures were obtained before and after each object integration. The
value of the seeing FWHM during the observing nights was between
0\farcs8 and 1\farcs2 as measured by the La Silla Differential Image
Motion Monitor. Using standard MIDAS routines the spectra were
bias subtracted, flat-field corrected, cleaned for cosmic rays and
wavelength calibrated. Cosmic rays were identified by comparing the
counts in each pixel with the local mean and standard deviation, and
then corrected by substituting a suitable value.  The instrumental
resolution was derived by measuring the FWHM of $\sim 30$ single
emission lines distributed all over the spectral range of a calibrated
comparison spectrum. It corresponds to a FWHM = $0.233\pm0.017$ \AA\
(i.e. $\sim11$ \kms\ at \ha). The single spectra of the same object were
aligned and coadded using their stellar-continuum centers as reference.
In each spectrum the center of the galaxy was defined by the center of a
Gaussian fit to the radial profile of the stellar continuum. The
contribution of the sky was determined from the edges of the resulting
frame and then subtracted. To study the \ha\ emission line, we
subtracted the underlying stellar continuum which was determined by
averaging a 2.5 \AA\ wide region with high S/N adjacent to the \ha\
line. A constant stellar continuum provides a good match to the
underlying distribution within this narrow wavelength range. For the
purpose of this paper the subtraction of a continuum with the proper
\ha\ absorption is not crucial since the same bidimensional shape is
observed also in the \nii\ $\lambda$6583 line. We did not model this
line bacause it falls at the edge of the sensitivity curve.

In addition to the spectroscopic material narrow-band \ha\ imaging of
NGC~2179 was performed on March 9-11, 1997 at the 1.8-m Vatican Advanced
Technology Telescope. A back illuminated 2048$\times$2048 Loral CCD with
$15\times15$ $\mu$m$^2$ pixels was used as detector at the aplanatic
Gregorian focus, f/9. It yielded a field of view of
$6\farcm4\times6\farcm4$ with an image scale of $0\farcs4$ pixel$^{-1}$
after a $2\times2$ on-line pixel binning.  We obtained $3\times10$
minutes emission-band images and $3\times2$ minutes Cousins $R-$band
images. The emission-band images were taken with an interference filter
($\lambda_c = 6630$ \AA; $\Delta\lambda_{\rm FWHM} = 70$ \AA) isolating
the spectral region characterized by the redshifted \ha\ and \nii\
$\lambda\lambda$6548, 6583 emission lines.  The data reduction was
routine.  Gaussian fit to field stars in the two final processed images
yielded point spread function FWHM of $1\farcs0$.  The continuum-free
image of NGC~2179 showing the galaxy \ha$+$\nii\ emission was obtained
by subtracting the $R$-band image, suitably scaled, from the
emission-band image.

\section{Results\label{results}}

The complex bidimensional structure (in the velocity-position map) of
the \ha\ emission line in NGC~2179 is shown in Fig.~1a. In the central
region ($r\la2''$) the line is highly tilted and extends up to $\pm250$
\kms.  The intensity distribution along the line shows two symmetric
peaks and a central minimum at $r=0$, $v=0$.  The shape of the emission
is such that at $r=\pm3''$ we observe $\Delta v = 200$ \kms.  Proceeding
further away from the center, $\Delta v$ increases up to an almost
constant value of $400$ \kms.

In the following we demonstrate that the peculiar shape and intensity
distribution of these emission lines is not produced by two
kinematically distinct components as the appearance could suggest, but
they are due to a unique velocity field traced by a thin gaseous disk
rotating in the combined potential of a central point-like mass
($v\rightarrow\infty$ as $r\rightarrow0$) embedded in an extended
stellar disk and to the deterioration caused by the instrumental
resolution and seeing. The presence of a gaseous disk in the principal
plane of NGC~2179 is supported by our \ha\ image, which shows a smooth
central structure characterized by the same ellipticity ($\epsilon
\simeq 0.2$) and major-axis position angle (P.A. $\simeq$ 170\degr) as
those of the stellar content observed in the $R-$band.  The bright knots
present in the observed spectrum are due to spiral arms which are also
visible in our \ha\ image.

We assume that the gas resides in an infinitesimally thin disk whose
mean motion is characterized by circular orbits in the plane of the
galaxy. At each position $(x,y)$ on the sky the line-of-sight velocity
profile is a Gaussian $\phi$ with mean $V(x,y)=V_c(R)\; x \sin i / R$
and dispersion $\sigma(R)$, where $R^2=x^2+(y/\cos i)^2$ is the radius
in the disk and $i$ (Table~1) the inclination of the galaxy. The
dispersion is given by $\sigma^2(R)=\sigma^2_{gas}(R)+\sigma^2_{instr}$,
where $\sigma_{gas}(R)$ is the intrinsic velocity dispersion of the gas
and $\sigma_{instr}$ is the instrumental dispersion (assuming a Gaussian
instrumental broadening function). The gas dispersion is assumed to be
isotropic and has been parametrized through
$\sigma_{gas}(R)=\sigma_0+\sigma_1 \exp(-R/R_t)$, where $R_t$ is the
scale length of the turbulence.

The circular velocity $V_c(R)$ is produced by the combined potential of
a point-like mass $M_\bullet$ and of the disk-like stellar component.
The contribution to the velocity due to the point-like mass is given by
$V_\bullet(R)=\sqrt{G M_\bullet/R}$. The contribution $V_\star(R)$ due
to the stars potential has been directly measured on the emission lines
of each spectrum at distances where both the seeing effect and the
point-like mass attraction are negligible, and it has been linearly
interpolated for smaller $R$ by imposing that $V_\star(0)=0$.  The
linearity of this interpolation is not crucial for the model since in
the inner regions the potential is dominated by the contribution of the
central point-like mass concentration. The resulting intrinsic velocity
profile of the disk is then computed as
$V_c^2(R)=V_\bullet^2(R)+V_\star^2(R)$.

The bidimensional model of the emission lines is given by
\begin{equation}
\Phi(v,S)=\int_{S-\frac{\Delta s}{2}}^{S+\frac{\Delta s}{2}} ds
    \int_{B-\frac{h}{2}}^{B+\frac{h}{2}} db
    \int\!\!\!\int_{-\infty}^{+\infty} ds' db' 
    \phi[v-V(s',b')] I(s',b') P(s'-s,b'-b)
\end{equation}
where $(S,B)$ are the coordinate along the slit and perpendicularly to
it respectively, while $h$ is the slit width and $\Delta s$ is the pixel
size of the detector. $I(s',b')$ is the intrinsic surface brightness
distribution of the disk and has been parametrized as $I(R)=I_0+I_1
\exp(-R/R_I)$, where $I_0$, $I_1$ and $R_I$ are free parameters.
$P(s'-s,b'-b)$ is the PSF which has been modeled as a Gaussian, owing to
the lack of a specific PSF image obtained at the time of the
observations. Note that the parameterizations for $I(R)$ and
$\sigma_{gas}(R)$ have been chosen because they are able to adequately
reproduce our data, but they have no further physical significance.

The line profile $\Phi(v,S)$, rebinned on a grid with steps $\Delta v$
(reciprocal dispersion) and $\Delta s$, can be directly compared to the
star-light-subtracted bidimensional spectrum obtained on the CCD. Using
the above model and the distance given in Table~1 we have obtained for
NGC~2179 the simulation of the \ha\ emission line shown in Fig.~1c,
which is remarkable for its similarity with the observed one (Fig.~1a).
The bright knots in the spectrum due to the spiral arms have not been
reproduced in our model.  The point-like mass of the best-fit model is
$1\times10^9$ M$_\odot$. It should be clear that our model shows only
the consistency of the observations with the presence of a central
point-like mass (black hole). However a central stellar density cusp
could also fit the data. In order to estimate the errors in the central
mass determination we have modeled the shape of the line for different
central point-like masses. In Fig.~1b and Fig.~1d we show two extreme
cases where the central mass is smaller and larger respectively by a
factor 5 than the mass of the best fit. On the basis of visual
comparison with the observations of a series of models we estimate that
our error could not be larger than a factor 3 ($\log M_\bullet = 9.0 \pm
0.5$ in solar units).

The bidimensional shape of the \ha\ emission line in the second galaxy
we observed, NGC~5064, is shown in Fig.~2a. Contrary to NGC~2179, it
does not present any peculiar central structure.  The line gives rise to
a standard rotation curve with the inner rigid-body rotation extended up
to $r=\pm4''$, and followed by a flat portion ($\Delta v = 400$ \kms).
By applying the same modeling technique as we did for NGC~2179 we
computed the shape of the emission lines as a function of decreasing
central mass until the complex structure described above tend to
disappear. The limiting case corresponds to a central point-like mass of
$5\times10^7$ M$_\odot$ and is illustrated in Fig~2b. Comparing this
model with the observed \ha\ line in NGC~5064 we deduce that in this
galaxy either the central mass is lower than $5\times10^7$ M$_\odot$ or
the unresolved Keplerian part of the gaseous disk does not give a
detectable contribution.

Emission lines with a bidimensional shape and an intensity distribution
similar to those observed in NGC~2179 have been observed also by Rubin
et al.\ (1997). We have selected in their sample the most representative
ones, namely NGC~4343 (Sa), NGC~4435 (S0) and NGC~4459 (S0). We applied
our modeling techniques to reproduce their isophotal maps, taking into
account their instrumental setup and seeing condition. The comparison
between the observations and our models is shown in Fig.~3, and it
appears quite satisfactory. In all the three cases the central tilted
part of the emission is characterized by the absence of a central
intensity peak, both in the model and in the observed line. The absence
of the flat region of the rotation curve produced by the potential of
the stellar disk in the case of NGC~4435 can be easily modeled by
adopting $I_0\simeq0$. The values of the central masses are of the same
order as that of NGC~2179 and are given in Table~1. A comparison between
a set of model lines obtained with different central masses and the
observations has been carried out for these galaxies, in the same way as
we did for NGC~2179. We estimate that the masses given in Table~1 for
the Rubin et al.\ (1997) galaxies are affected by uncertainties of the
same order of magnitude as in NGC~2179.

\section{Discussion}

In the previous paragraphs we have shown that the peculiar bidimensional
shape of the emission  lines in a sample of four galaxies is consistent
with the effect produced by the combined potential of a central
point-like mass and of an extended component. In this way we were able
to point out the presence of central mass concentrations of the order of
$10^9$ M$_\odot$ in a sample of disk galaxies, observed with
ground-based telescopes.

With the observations presented in this paper we demonstrate the
possibility of detecting central mass concentrations in galaxies with
ground-based telescopes properly equipped using the CNKDs as probes. The
masses which can be detected in this way are larger than $5\times10^7$
M$_\odot$ at the distance of the Virgo cluster. Up to now the higher
resolution offered by HST has not contributed to the detection, using
CNKDs, of central masses lower than this limit. In fact the four
galaxies so far studied with HST posses central masses of the same order
of ours. A simulation similar to the one used in this paper predicts
that HST, equipped with the Space Telescope Imaging Spectrograph will
allow to detect central masses down to the level of $5\times10^6$
M$_\odot$. Note that although VLBI spectroscopy of H$_2$O masers
delivers much higher angular resolution, this technique is limited by
the availability of suitably bright sources. We are inclined to think
that the detection of lower mass black holes will constitute one of the
most proper use of HST, which also allows to put more stringent
constraint on the size of the region containing the central mass.

For the four galaxies of this paper with positive detection we can
summarize our measurements with a single median value of the ratio of
the central mass to the luminosity of the bulge component
$M_\bullet/L_{\it B,\, bulge}\sim0.16$. This value is one order of
magnitude larger than the median value derived by Ho (1998) from a
sample of 20 objects but still within the scatter. If we assume for
$M_\bullet$ of NGC~5064 the upper limit we derived, also this galaxy
falls within the scatter of the relation $M_\bullet-L_{\it B,\, bulge}$
(Ho 1998).

\acknowledgments

We thank Dave Burstein and Vera Rubin for helpful discussions. JCVB
acknowledges the support by a grant of the Telescopio Nazionale Galileo
and Osservatorio Astronomico di Padova.

\clearpage

\begin{deluxetable}{llllrlrlrll}
\scriptsize
\tablecaption{Parameters of the modeled galaxies.}
\tablehead{
\colhead{object} & \colhead{type} & \colhead{type} & \colhead{$B^0_T$} &
\colhead{P.A.} & \colhead{$i$} & \colhead{$V_{\odot}$} & \colhead{$D$} &
\colhead{scale} & \colhead{$M_{B,{\it bulge}}$} & \colhead{$M_\bullet$}\\
\colhead{[name]} & \colhead{[RSA]} & \colhead{[RC3]} & \colhead{[mag]} &
\colhead{[\degr]} & \colhead{[\degr]} & \colhead{[\kms]} &
\colhead{[Mpc]} & \colhead{[pc$''^{-1}$]} & \colhead{[mag]} &
\colhead{[$10^{8}$M$_\odot$]}\\
\colhead{(1)} & \colhead{(2)}  & \colhead{(3)} & \colhead{(4)}  &
\colhead{(5)} & \colhead{(6)}  & \colhead{(7)} & \colhead{(8)}  &
\colhead{(9)} & \colhead{(10)} & \colhead{(11)}
}
\startdata
NGC~2179\tablenotemark{a} & Sa & SAS0 & 12.83 & 170 & 51 & 2885$\pm$15 &
35.6 & 172.6 & -19.07 & 10 \nl
NGC~5064\tablenotemark{a} & Sa & PSA2$\ast$ & 11.67 & 38 & 65 &
2980$\pm$15 & 36.7 & 177.9 & -19.78 & $<$ 0.5 \nl
NGC~4343\tablenotemark{b} & \nodata & SAT3$\ast$ & 12.37 & 133 & 78 &
1002$\pm$10 & 17.0 & 82.4 & -17.24 & 5 \nl
NGC~4435\tablenotemark{b} & SB0$_1(7)$ & LBS0 & 11.61 & 13 & 90 &
806$\pm$10 & 17.0 & 82.4 & -18.93 & 10 \nl
NGC~4459\tablenotemark{b} & S0$_3$(3) & LAR$+$ & 11.21 & 110 & 42 &
1183$\pm$10 & 17.0 & 82.4 & -19.20 & 10 \nl
\enddata
\tablenotetext{a}{From our sample}
\tablenotetext{b}{from the sample of Rubin et al.\ (1997)}
\tablecomments{Col.~(2): morphological classification from Sandage \&
    Tamman (1981). Col.~(3): morphological classification from de
    Vaucouleurs et al.\ (1991, hereafter RC3). Col.~(4): total $B$
    magnitude after correcting for extinction and redshift from RC3.
    Col.~(5): major-axis position angle from RC3. Col.~(6): inclination
    from Rubin et al.\ (1997) except for NGC~2179 and NGC~5064 (Tully
    1988). Col.~(7): heliocentric systemic velocity derived as the center
    of symmetry of the gas rotation curve. For NGC~2179, NGC~5064 it is
    taken from our data, and for NGC~4343, NGC~4435, NGC~4459 it is
    taken from Rubin et al.\ (1997). Col.~(8): distance of NGC~2179 and
    NGC~5064 is derived from the heliocentric velocity corrected for the
    motion of the Sun with respect of the Local Group by $\Delta
    V=300\cos{b}\sin{l}$ with $H_0=75$ \kms\ Mpc$^{-1}$. The galaxies of
    the Rubin's sample are members of the Virgo cluster and we assumed a
    distance of 17 Mpc (Freedman et al.\ 1994). Col.~(10): statistical
    estimate of the absolute $B$ magnitude of the bulge derived
    following Simien \& de Vaucouleurs (1986). Col.~(11): value of the
    central mass concentration derived as described in
    Sect.~\ref{results}}.
\end{deluxetable}

\clearpage
\vspace*{3truecm}
\epsscale{1}
\plotone{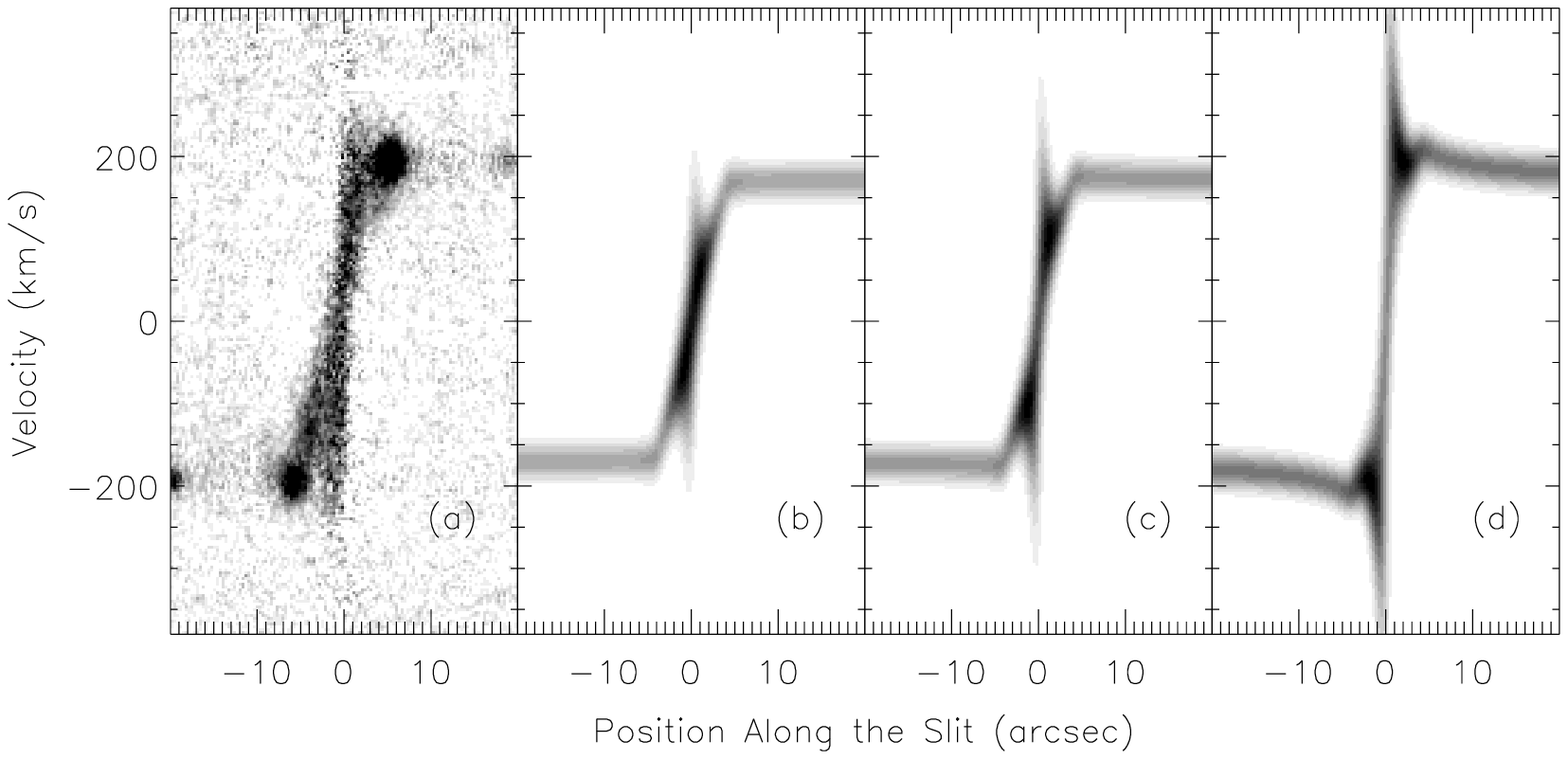}
\figcaption[]{{\it (a):\/} The \ha\ emission line, observed along the
    major axis of NGC~2179, after subtraction of the stellar continuum.
    {\it (b), (c) and (d):\/} Models of the NGC~2179 \ha\ line [shown in
    the same scale of {\it panel (a)\/}] obtained with different
    point-like central masses.  {\it (b):\/} $M_\bullet=2\times10^8$
    M$_\odot$.  {\it (c):\/} $M_\bullet=1\times10^9$ M$_\odot$
    corresponding to our best-fit model.  {\it (d):\/}
    $M_\bullet=5\times10^9$ M$_\odot$.}

\clearpage
\vspace*{3truecm}
\epsscale{.55}
\plotone{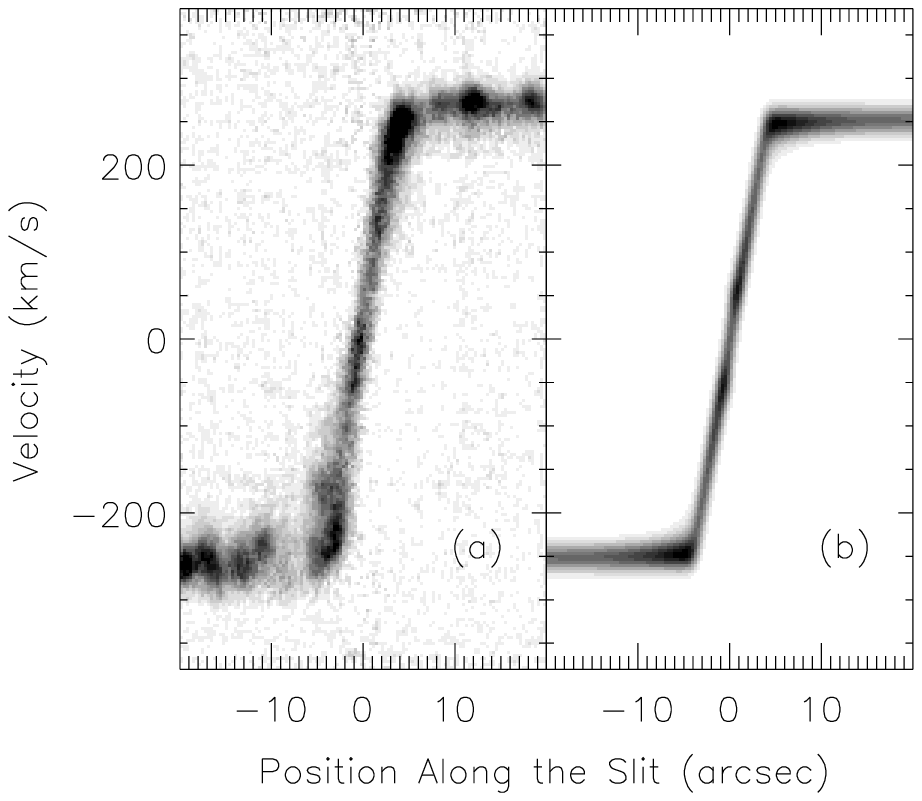}
\figcaption[]{{\it (a):\/} The \ha\ emission line, observed along the
    major axis of NGC~5064, after subtraction of the stellar continuum.
    {\it (b):\/} Model of the NGC~5064 \ha\ line obtained with the
    highest point-like central mass which can be added without
    significantly disturbing the general shape of the line
    ($M_\bullet=5\times10^7$ \msun).}

\clearpage
\epsscale{.8}
\plotone{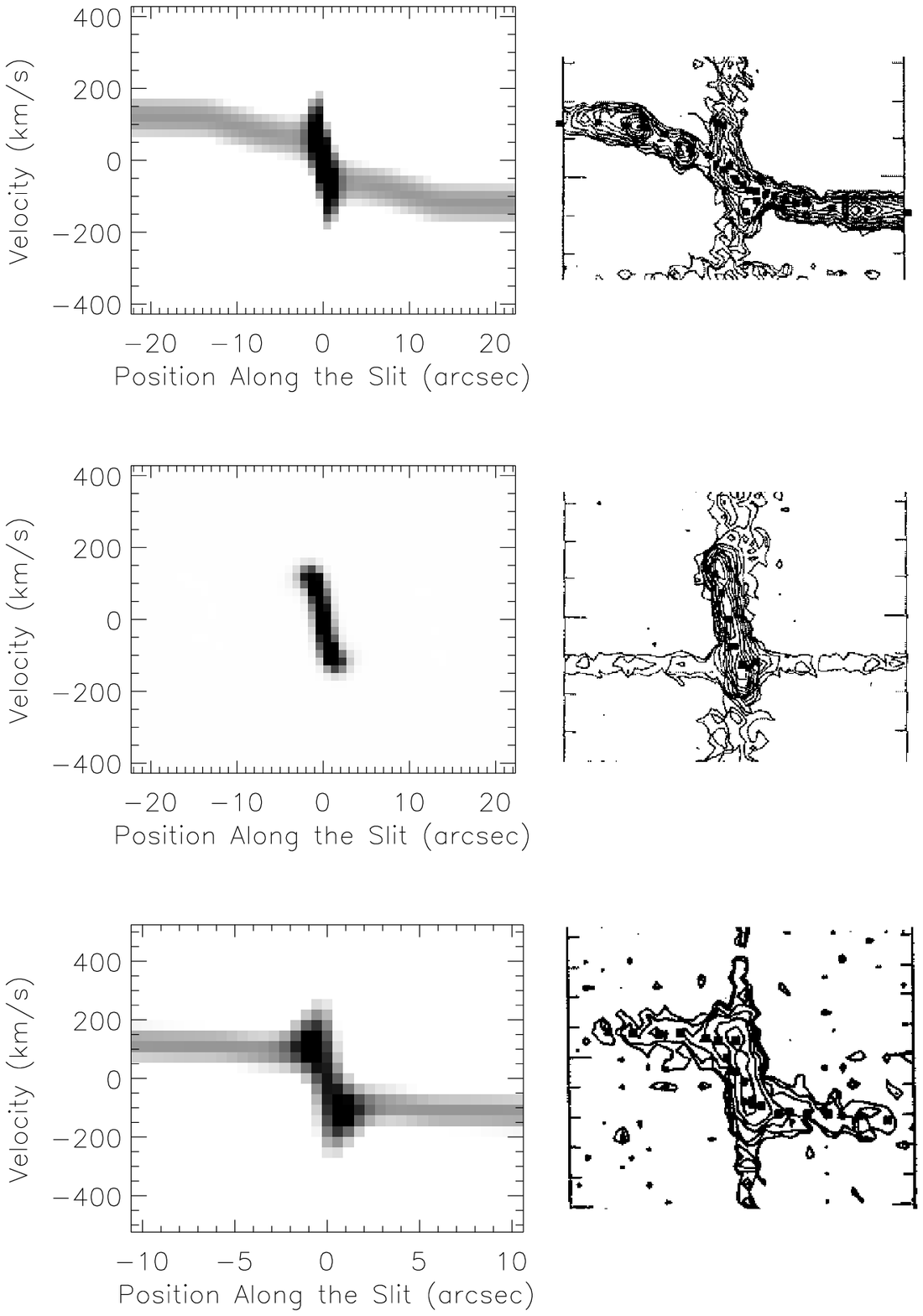}
\figcaption[]{Best-fit models ({\it left column\/}) of the \ha\ emission
    lines observed (their isophotal maps with the same scale are shown
    in the {\it right column\/}) by Rubin et al.\ (1997) along the major
    axis of NGC~4343 ({\it top panels\/}), NGC~4435 ({\it middle
    panels\/}) and NGC~4459 ({\it bottom panels\/}).}

\end{document}